\documentclass[twocolumn,aps,showpacs,longtable]{revtex4}
\usepackage{amsfonts,amsmath,graphicx,latexsym}
\usepackage{epsfig}
\newcommand{\vc}[1]{\mathbf{#1}}

\begin{document}
\title{Bose Condensed Gas in Strong Disorder Potential With Arbitrary 
Correlation Length}
\author{Patrick Navez$^{1,2}$, Axel Pelster$^{1}$, Robert Graham$^{1}$}
\affiliation{
$^{1}$Universitaet Duisburg-Essen,
Universitaet Duisburg-Essen, Lotharstrasse 1, 47048 Duisburg, 
Germany, \\ 
$^{2}$Labo Vaste-Stoffysica en Magnetisme, 
Katholieke Universiteit Leuven,
Celestijnenlaan 200 D,
B-3001 Heverlee, Belgium}

\date{\today}
\begin{abstract}
We study the properties of a dilute 
Bose condensed gas at zero temperature 
in the presence of a strong random potential 
with arbitrary correlation length. Starting from the underlying
Gross-Pitaevskii equation, we use the random phase approximation in order 
to get a closed integral equation for the averaged density 
distribution which allows the determination of 
the condensate and the superfluid density. 
The obtained results generalize those of 
Huang and Meng (HM) to strong disorder. In particular, 
we find the critical value of the disorder strength, where
the superfluid phase disappears by a first-order phase transition. 
We show how this critical value changes 
as a function of the correlation length. 
\end{abstract}
\maketitle
\section{Introduction}
An ultracold atom gas in the presence of disordered environments is becoming 
a subject of increasing experimental and theoretical research
activities. Generally, one would like to understand 
how the condensation and the 
superfluid properties of ultracold gases are influenced 
by a spatially random force on the atoms.
In some experiments,  
the random potential is created by optical means to show 
its effects on the transport 
properties of a Bose gas \cite{Aspect,Inguscio,Ertmer}. 
In most others, however, one  
must rather face the reality of
unavoidable external random forces which are 
induced either by the roughness of a 
dielectric surface \cite{Perrin}, by the magnetic field 
along wires with current
irregularities \cite{Schmiedmayer}, or by different localized atomic species
\cite{Sengstock}. Furthermore, recent theoretical results on 
the impact of randomness on bosons in lattices
are reviewed in Refs.~\cite{Krutitsky,Lewenstein}.

Most theoretical 
studies on a 3D disordered Bose gas are limited to calculations up to the second-order
in the random potential.  
In these works, the weak disorder induces only small corrections to the
condensate depletion, the superfluid density \cite{HM}, 
the collective 
excitations and their damping \cite{Giorgini}, 
and the condensation critical point 
\cite{Lopatin}. The extension to strong disorder has so far been analysed only  
numerically in Ref.~\cite{Giorgini2}. 
More recently, an analytical mean-field study \cite{Pelster}, which takes into 
account higher order corrections, 
has shown the possibility 
of having a transition from a superfluid phase to a Bose glass phase where 
the spatial long-range correlations have completely disappeared. 

In this paper, we address the issue of the influence of strong disorder 
at zero temperature for a finite correlation length \cite{Kobayashi,Timmer}.
The random potential follows a Gaussian distribution and is said to be 
uncorrelated in the case where all Fourier components contribute 
equally to the randomness, 
while it is correlated when the influence of the Fourier components 
falls off for wavenumbers larger the inverse correlation length $\xi$.
As encoutered in experiments \cite{Schmiedmayer}, 
we choose a Lorentzian correlation 
function. Usually, this length appears to be much bigger 
than the healing length and thus affects the condensate properties.
  
In order to simplify this physical problem, we assume that all particles 
occupy the same quantum state, for which the macroscopic wave-function obeys 
the Gross-Pitaevskii equation in the presence of an external spatially 
random force. 
In order to solve this stochastic nonlinear differential equations,
we apply the random phase approximation (RPA) \cite{Pines} and take  
the ensemble average over all possible realisations of the associated 
potential. In the clean case without a random potential, 
this gapless and conserving approximation has been successfully 
used in the context of calculating  
of the collective excitations at finite temperatures 
\cite{Reidl,Fliesser} and in kinetic theory \cite{condenson}. 
In our case with disorder, we obtain the particle density distribution 
beyond the lowest order expansion in the random potential. 
With this we extend
the seminal work of Ref.~\cite{HM} to strong disorder, which has
the consequence 
that the superfluid phase disappears 
by a first-order transition. We show how the critical value 
of the disorder strength, for which this transition   
occurs, depends on the correlation length. 
\section{Momentum Distribution}
We start from the Gross-Pitaevskii equation at zero temperature written in 
Fourier space. Defining the Fourier components according to
$\psi(\vc{r})=\sum_{\vc{k}}e^{i\vc{k}.\vc{r}}\alpha_{\vc{k}}/\sqrt{V}$,
we get in units with $\hbar =1$:
\begin {eqnarray}
\label{EVOLV}
\left(i\frac{\partial}{\partial t}-\frac{\vc{k}^2}{2m}\right)\alpha_\vc{k}&=&
\sum_{\vc{q}}U_{\vc{q}}\alpha_{\vc{k}-\vc{q}}\nonumber \\
&& + \frac{g}{V}\sum_{\vc{k'},\vc{q}}\alpha^*_{\vc{k'}}
\alpha_{\vc{k}+\vc{q}}\alpha_{\vc{k'}-\vc{q}} \, ,
\end{eqnarray}
where the random potential $U_{\vc{q}}$ follows a Gaussian distribution 
$\langle U_{\vc{q}}U_{\vc{q'}}\rangle=\delta_{\vc{q},-\vc{q'}}R(\vc{q})/V$.
The quadratic amplitude is assumed to be Lorentzian $R(\vc{q})=R/(1+\xi^2 \vc{q}^2)$, 
i.e.~it is correlated below the wavenumber $1/\xi$.
We assume that a macroscopic fraction of the 
condensate moves with a velocity $\vc{k_s}/m$.

In order to derive a dynamic dielectric function for the total fluid,
we define the bilinear combination 
\begin{eqnarray}
\label{RHO}
\rho_{\vc{k},\vc{q}}=\alpha^*_{\vc{k}}\alpha_{\vc{k}+\vc{q}} \, .
\end{eqnarray}
It represents an excitation of momentum $\vc{q}$ created from
a particle which releases its momentum from $\vc{k}+\vc{q}$ to
$\vc{k}$. In particular, it allows the definition of
the Fourier component of the density fluctuation: 
$\rho_{\vc{q}}=\sum_{\vc{k}} \rho_{\vc{k},\vc{q}}=
\rho^*_{\vc{-q}}$. The dynamic evolution of (\ref{RHO}) 
following from (\ref{EVOLV}) is given by:
\begin{eqnarray}
\label{he}
\hspace*{-30mm}
i\frac{\partial}{\partial t}\rho_{\vc{k},\vc{q}}
&=&(\epsilon_{\vc{k}+\vc{q}}
-\epsilon_{\vc{k}})
\rho_{\vc{k},\vc{q}}+
 \\ \nonumber 
\sum_{\vc{q'}}
(U_{\vc{q'}}&+&\frac{g}{V} \sum_{\vc{k'}}\alpha_\vc{k'}^*\alpha_{\vc{k'+q'}})
(\alpha^*_{\vc{k}}\alpha_{\vc{k+q}-\vc{q'}}-
\alpha^*_{\vc{k}+\vc{q'}}\alpha_\vc{k+q}) \, .
\end{eqnarray}
The technical details for implementing the 
random phase approximation in the clean case can be found either in 
Ref.~\cite{Pines} or in Ref.~\cite{condenson} where 
a finite-temperature Bose gas has been considered with 
$\rho_{\vc{k},\vc{q}}$ being a  quantum operator. Here, we shall briefly 
repeat this procedure by
considering random variables instead of operators. 

We treat the quartic terms in (\ref{he}) 
within a factorisation procedure 
and use the property that 
$\langle \rho_{\vc{k},\vc{q}} \rangle =0$ for $\vc{q}\not=0$ in 
an homogeneous gas. 
For any quartic term, 
we approximate
\begin{eqnarray}\label{fact}
&& \hspace*{-10mm}\alpha^*_{\vc{k_1}}\alpha^*_{\vc{k_2}}
\alpha_{\vc{k_3}}\alpha_{\vc{k_4}}
- \langle \alpha^*_{\vc{k_1}}\alpha^*_{\vc{k_2}}
\alpha_{\vc{k_3}}\alpha_{\vc{k_4}} \rangle \simeq
\nonumber \\ &+& 
\langle |\alpha_{\vc{k_1}}|^2 \rangle \delta_{\vc{k_1},\vc{k_3}}
\alpha^*_{\vc{k_2}}\alpha_{\vc{k_4}}
\nonumber \\
&+&\langle |\alpha_{\vc{k_2}}|^2 \rangle 
\delta_{\vc{k_2},\vc{k_4}}
(1+\delta_{\vc{k_1},\vc{k_2}}\delta_{\vc{k_3},\vc{k_4}})
\alpha^*_{\vc{k_1}} \alpha_{\vc{k_3}}
\nonumber \\
&+&
\langle |\alpha_{\vc{k_1}}|^2 \rangle 
\delta_{\vc{k_1},\vc{k_4}}
(1-\delta_{\vc{k_1},\vc{k_2}}-\delta_{\vc{k_3},\vc{k_4}})
\alpha^*_{\vc{k_2}}\alpha_{\vc{k_3}}
\nonumber \\
&+&
\langle |\alpha_{\vc{k_2}}|^2 \rangle  
\delta_{\vc{k_2},\vc{k_3}}
(1-\delta_{\vc{k_1},\vc{k_2}}-\delta_{\vc{k_3},\vc{k_4}})
\alpha^*_{\vc{k_1}}\alpha_{\vc{k_4}} \,
\end{eqnarray}
which avoids double counting. For example,  
in case of $\vc{k_1}=\vc{k_2}$ and 
$\vc{k_3}\not=\vc{k_4}$
the approximation reduces to two terms only which is important for
contributions involving the macroscopic component $\vc{k_s}$ 
of the condensate.  Since the  
average over the 
quartic term in Eq.(\ref{fact}) applied in (\ref{he}) 
involves components with 
the total transfer momentum $\vc{q}\not=0$, 
it will not contribute for an homogeneous gas. 

Through this procedure, 
the RPA keeps among all terms those combinations 
involving products of off-diagonal terms $\rho_{\vc{k'},\vc{q}}$ 
and averaged diagonal ones $n_\vc{k''}=\langle |\alpha_{\vc{k''}}|^2 \rangle $ 
for all possible values of $\vc{k'}$
and $\vc{k''}$,
and neglects all others combinations.
Therefore, we remove contributions which are bilinear in
$\rho_{\vc{k'},\vc{q'}}$
for $\vc{q'}\not=\vc{q},\vc{0}$. 
In this way, we 
obtain the linear integral equation for $\vc{q} \not= \vc{0}$:
\begin{eqnarray}
\label{rho4}
\left[i\frac{\partial}{\partial t}
-(\epsilon_{\vc{k}+\vc{q}}-\epsilon_{\vc{k}})\right]
\rho_{\vc{k},\vc{q}}=
\nonumber \\
(U_{\vc{q}}+\frac{g \rho_{\vc{q}}}{V})
(n_{\vc{k}}-n_{\vc{k+q}})
+\frac{g\rho_{\vc{q}}}{V}( n'_{\vc{k}}-n'_{\vc{k+q}})
\nonumber \\
+ (1-\delta_{\vc{k},\vc{k_s}}-\delta_{\vc{k},\vc{k_s}-\vc{q}})
\!\!\!\! \sum_{\vc{k'}\not=
\vc{k_s},\vc{k_s}-\vc{q}}
\frac{g\rho_{\vc{k'},\vc{q}}}{V}
n_{\vc{k_s}}
\, .
\end{eqnarray}
Here $n'_{\vc{k}}=(1-\delta_{\vc{k_s},\vc{k}})n_{\vc{k}}$ refers 
to the disordered part of the condensate which consists of all its part 
which are not at the wavenumber $\vc{k_s}$.
Note at this stage that $\rho_{\vc{k},\vc{q}}$ is still 
a random variable. Thus, we should still take 
the ensemble average over 
any non-vanishing combination like 
$\langle U_{-\vc{q}} \rho_{\vc{k},\vc{q}} \rangle$
and $\langle \rho_{\vc{k'},-\vc{q}} \rho_{\vc{k},\vc{q}} \rangle$ 
and solve the resulting equations. Equivalently, here we directly solve 
Eq.~(\ref{rho4}) for $\rho_{\vc{k},\vc{q}}$ 
and perform the disorder average at a later stage.

In order to make the link with the dielectric formalism, let us 
assume for the moment that the potential has a temporal dependence 
of the form 
$U_{\vc{q}}(t)=\exp(-i\omega t)U_{\vc{q},\omega}$. Then the solution is 
of the form $\rho_{\vc{k},\vc{q}}(t)=
\exp(-i\omega t)\rho_{\vc{k},\vc{q},\omega}$ as well. 
Under these conditions, we find a solution: 
\begin{eqnarray}\label{sol}
\lefteqn{\rho_{\vc{k},\vc{q},\omega}=
\frac{
(U_{\vc{q},\omega}+ 2g \rho_{\vc{q},\omega}/V)
(n_{\vc{k}}-n_{\vc{k+q}})}{\omega+i0_+ -
(\epsilon_{\vc{k}+\vc{q}}-\epsilon_{\vc{k}})}}
\nonumber \\ \nonumber
&&\times
\frac{
(\omega+i0_+ -
\frac{\vc{k_s}.\vc{q}}{m})^2-(\frac{\vc{q}^2}{2m})^2
}
{(\omega+i0_+ -
\frac{\vc{k_s}.\vc{q}}{m})^2-(\frac{\vc{q}^2}{2m})^2+
\frac{g n_{\vc{k_s}}\vc{q}^2}{mV}
(\delta_{\vc{k},\vc{k_s}}+\delta_{\vc{k},\vc{k_s-q}})}
\nonumber \\
\end{eqnarray}
which is similar to Ref.~\cite{condenson}. 
The Fourier component of the density fluctuations can 
therefore be written in the form: 
\begin{eqnarray}\label{resp}
\rho_{\vc{q},\omega}= \chi(\vc{q},\omega)U_{\vc{q},\omega} \, ,
\end{eqnarray}
with the susceptibility 
\begin{eqnarray}\label{K}
\chi(\vc{q},\omega)=
\frac{V}{2g}\left[\frac{1}{{\cal K}(\vc{q},\omega)}-1\right]\, .
\end{eqnarray}
The dynamic dielectric function ${\cal K}(\vc{q},\omega)$ defined 
by (\ref{K}) for the 
total fluid can be decomposed into \cite{condenson}
\begin{eqnarray}
\label{kal}
{\cal K}(\vc{q},\omega)
&=&{\cal K}_n(\vc{q},\omega) \nonumber \\
&&\hspace*{-5mm}+\frac{-\frac{2g n_{\vc{k_s}}}{V}
\frac{\vc{q}^2}{m}}
{\left(\omega+i0_+ -
\frac{\vc{k_s}.\vc{q}}{m}\right)^2-\left(\frac{\vc{q}^2}{2m}\right)^2
+\frac{g n_{\vc{k_s}} \vc{q}^2}{mV}}\, ,
\end{eqnarray}
where we obtain for the disordered part of the fluid:
\begin{eqnarray}
\label{kn}
{\cal K}_n(\vc{q},\omega)=
1- \frac{2g}{V}\sum_{\vc{k}}
\frac{n'_{\vc{k}}-n'_{\vc{k+q}}}{
\omega+i0_+ -
\frac{\vc{k}.\vc{q}}{m}-\frac{\vc{q}^2}{2m}} \, ,
\end{eqnarray}
(cf. also \cite{condenson,Graham}). 
In this way, we recover the same expression as in Ref.~\cite{Fliesser,
Graham} for 
the susceptibility function $\chi(\vc{q},\omega)$.  

In the special case of a time-independent potential 
$U_{\vc{q},\omega}=U_{\vc{q}}\delta_{\omega,0}$ which is of interest here, the
solution (\ref{sol}) can be used to define a self-consistent 
relation that allows the calculation of $n'_{\vc{k}}$. At first, combining 
Eqs.~(\ref{sol})--(\ref{kn}), we get the response function 
for a particle of the condensate to be excited with momentum $\vc{q}$:
\begin{eqnarray}
\label{respc}
\rho_{\vc{k_s},\vc{q}}=
n_\vc{k_s}\frac{U_{\vc{q}}}{\left(i0_+ -
\frac{\vc{k_s}.\vc{q}}{m}-
\frac{\vc{q}^2}{2m}\right){\tilde{{\cal K}}}(\vc{q})}\, .
\end{eqnarray}
For $\omega=0$, the screening factor ${\tilde{\cal{K}}}(\vc{q})$ 
defined by (\ref{respc}) for any external force 
acting on the condensate particles is:
\begin{eqnarray}\label{kalc}
&&\hspace*{-9mm}{\tilde{{\cal K}}}(\vc{q})=
\nonumber \\
&&\hspace*{-9mm}\frac{-(\frac{\vc{k_s}.\vc{q}}{m})^2+(\frac{\vc{q}^2}{2m})^2}
{{\cal K}_n(\vc{q},0)\left[
{\epsilon^B_{\vc{q}}}^2-( \frac{\vc{k_s}.\vc{q}}{m})^2\right]
- \left[{\cal K}_n(\vc{q},0)-1 \right]\frac{2g n_{\vc{k_s}}\vc{q}^2}{mV}}
\, .
\end{eqnarray}
In the simple special  case $n'_\vc{k}=0$, i.e. ${\cal K}_n(\vc{q},0)=1$ and 
$\vc{k_s}=0$, we obtain more simply:
\begin{eqnarray}
\label{resp1}
{\tilde{{\cal K}}}(\vc{q})=\frac{1}{1+(4mgn_0/V)/\vc{q}^2} \, .
\end{eqnarray}
This formula shows that the random force acting on the condensate particles 
is screened 
for momenta below the inverse of the healing length which plays here 
the role of the Debye length for the condensate.  

Let us now notice that:
$\langle \rho^*_{\vc{k_s},\vc{q}} \rho_{\vc{k_s},\vc{q}} \rangle 
=\langle |\alpha_{\vc{k_s}}|^2 |\alpha_{\vc{k_s +q}}|^2 \rangle
= n_{\vc{k_s}}n'_{\vc{k_s}+\vc{q}}$ for $\vc{q}\not= 0$ whereas the 
fluctuations of the macroscopic $\vc{k_s}$ component around 
the average are small in the thermodynamic limit.
Using this relation in (\ref{respc}) and (\ref{kalc}), 
we arrive at the following non-linear integral 
equation for $n'_{\vc{k_s+q}}$:
\begin{eqnarray}
\label{eqn}
&& \hspace*{-11mm}n'_{\vc{k_s}+\vc{q}}=
\nonumber \\
&& \hspace*{-11mm}\frac{R(\vc{q})(\epsilon_{\vc{k_s-q}}-\epsilon_{\vc{k_s}})^2 n_\vc{k_s}/V}
{|{\cal K}_n(\vc{q},0)[
{\epsilon^B_{\vc{q}}}^2-(
\frac{\vc{k_s}.\vc{q}}{m})^2]
-
({\cal K}_n(\vc{q},0)-1)\frac{2g n_{\vc{k_s}}}{V}
\frac{\vc{q}^2}{m}|^2} \, ,
\end{eqnarray}
where $\epsilon^B_{\vc{q}}=\sqrt{c_B^2 \vc{q}^2 +
(\vc{q}^2 / 2m})^2$
denotes the Bogoliubov excitation energy and 
$c_B=\sqrt{g n_{\vc{k_s}}/ m V}$ is the sound 
velocity in the absence of disorder. 
The nonlinearity comes from the fact that the dieletric function 
${\cal K}_n(\vc{q},0)$
depends  via Eq.~(\ref{kn}) on $n'_{\vc{k_s+q}}$.  
Let us first consider the case of weak disorder where we can approximate 
${\cal K}_n(\vc{q},0) \simeq 1$ and recover 
the second-order result for the disordered components of the condensate 
\cite{HM}:
\begin{eqnarray}
n'_{\vc{k_s}+\vc{q}}=
\frac{R(\vc{q})(\epsilon_{\vc{k_s-q}}-\epsilon_{\vc{k_s}})^2 n_\vc{k_s}/V}
{\left[
{\epsilon^B_{\vc{q}}}^2-\left(\frac{\vc{k_s}.\vc{q}}{m}\right)^2\right]^2} \, .
\end{eqnarray}
This formula is regular provided the Landau stability criterion 
$\epsilon^B_{\vc{q}}>|\vc{k_s}.\vc{q} / m|$ is  satisfied. 
An increase of $\vc{k_s}$ would 
increase the disordered part of the fluid until 
a singularity is reached leading to an instability.    
For $\vc{k_s}=0$, we recover the weak-disorder result 
for this part  \cite{HM}:
\begin{eqnarray}
\sum_\vc{q} n'_{\vc{q}}= \sum_\vc{q}
\frac{R(\vc{q})n_{\vc{0}}/V}
{[
\epsilon_{\vc{q}}^2+2gn_{\vc{0}}/V]^2}
=R \frac{m^{3/2}}{4\pi^2}\sqrt{\frac{V}{gn_0}} n_0 \, .
\end{eqnarray}
The superfluid fraction $n_s$ is defined as the part of the fluid which 
is moving at a velocity $\vc{k_s}/m$ and is found by calculating the total 
momentum of the gas:
\begin{eqnarray}
\vc{P}=\sum_{\vc{k}} \vc{k} n_{\vc{k}}
=\vc{k_s} n_s \, .
\end{eqnarray}
In the limit of small velocity $\vc{k_s}/m$, we get the expression:
\begin{eqnarray}
n-n_s=\frac{1}{3}\sum_\vc{k} \left(\vc{k}.\frac{\partial}{\partial \vc{k_s}}
n_{\vc{k_s+k}}\right)\biggr|_{\vc{k_s}=0} \, .
\end{eqnarray}
Noticing that ${\cal K}_n(\vc{q},0)$ with the solution $n'_{\vc{k}}$  
is an even function of  $\vc{k_s}$, we deduce from Eq.~(\ref{eqn}): 
\begin{eqnarray}\label{hms}
n-n_s=\frac{4}{3}(n - n_0) \, .
\end{eqnarray}
This relation is identical to the one in Ref.~\cite{HM} 
but is here found to remain valid in the RPA for the more 
general case of strong disorder. 

\begin{figure}[t]
\scalebox{0.5}{
\includegraphics{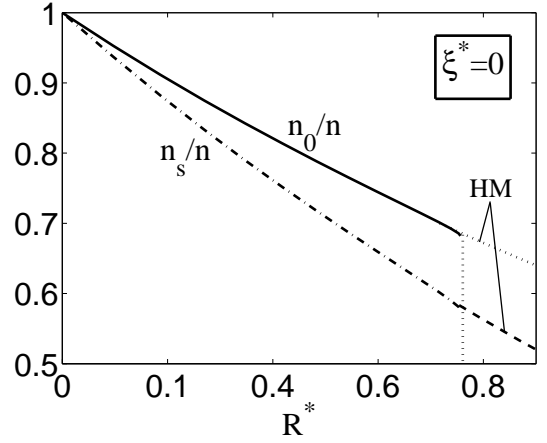}}
\caption{Clean part of the condensate fraction $n_\vc{0}$ 
as a function of $R^*$ in the case of uncorrelated disorder in the RPA model 
(full curve) and in the HM model (dotted curve) and the corresponding 
superfluid fraction $n_s$
(dot-dashed curve in RPA and dashed curve in HM model).
}
\label{fig1}
\end{figure}
 
\begin{figure}[t]
\scalebox{0.5}{
\includegraphics{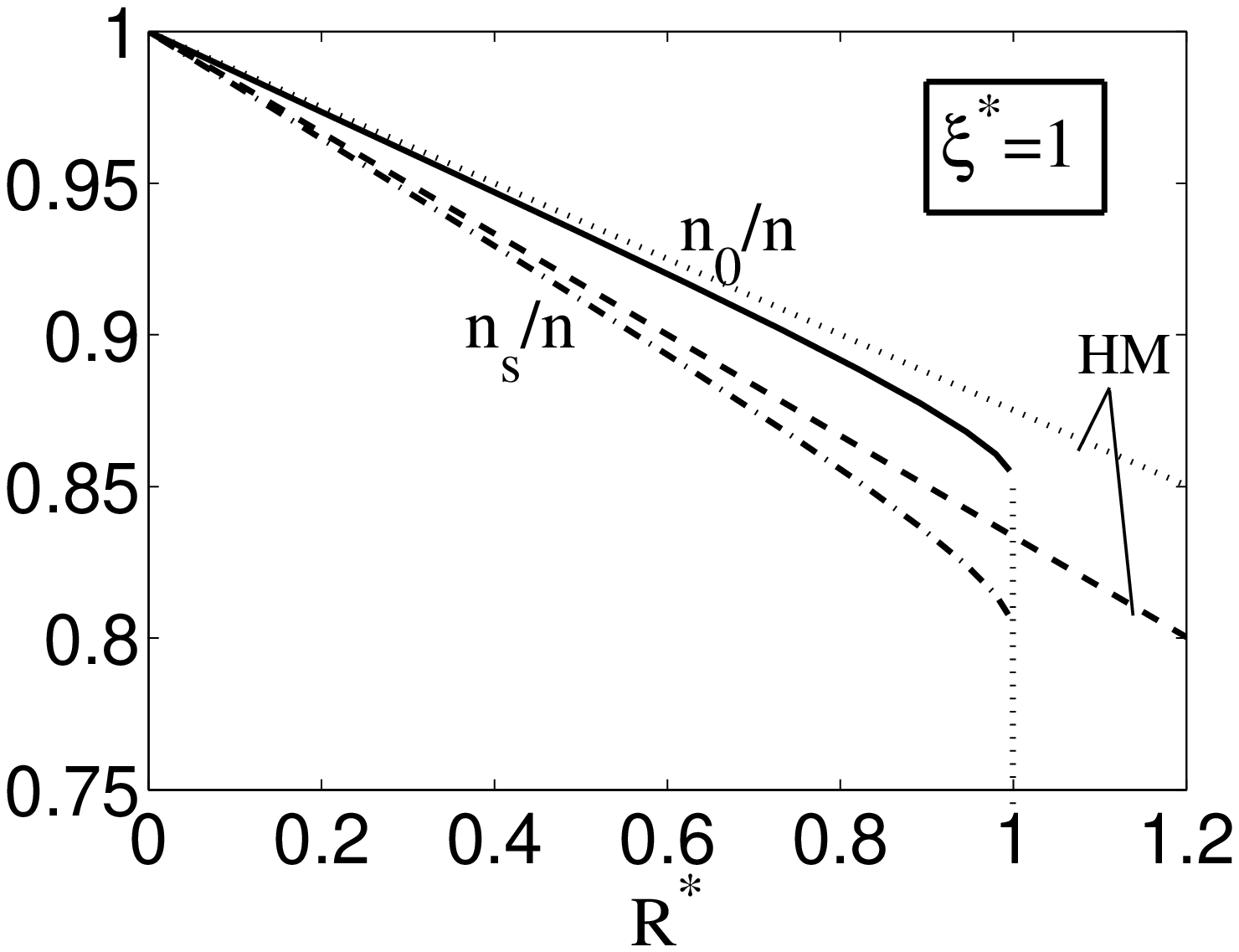}}
\caption{Same as for Fig.1 but for correlated disorder with $\xi^*=1$.
}
\label{fig2}
\end{figure}

\begin{figure}[t]
\scalebox{0.5}{
\includegraphics{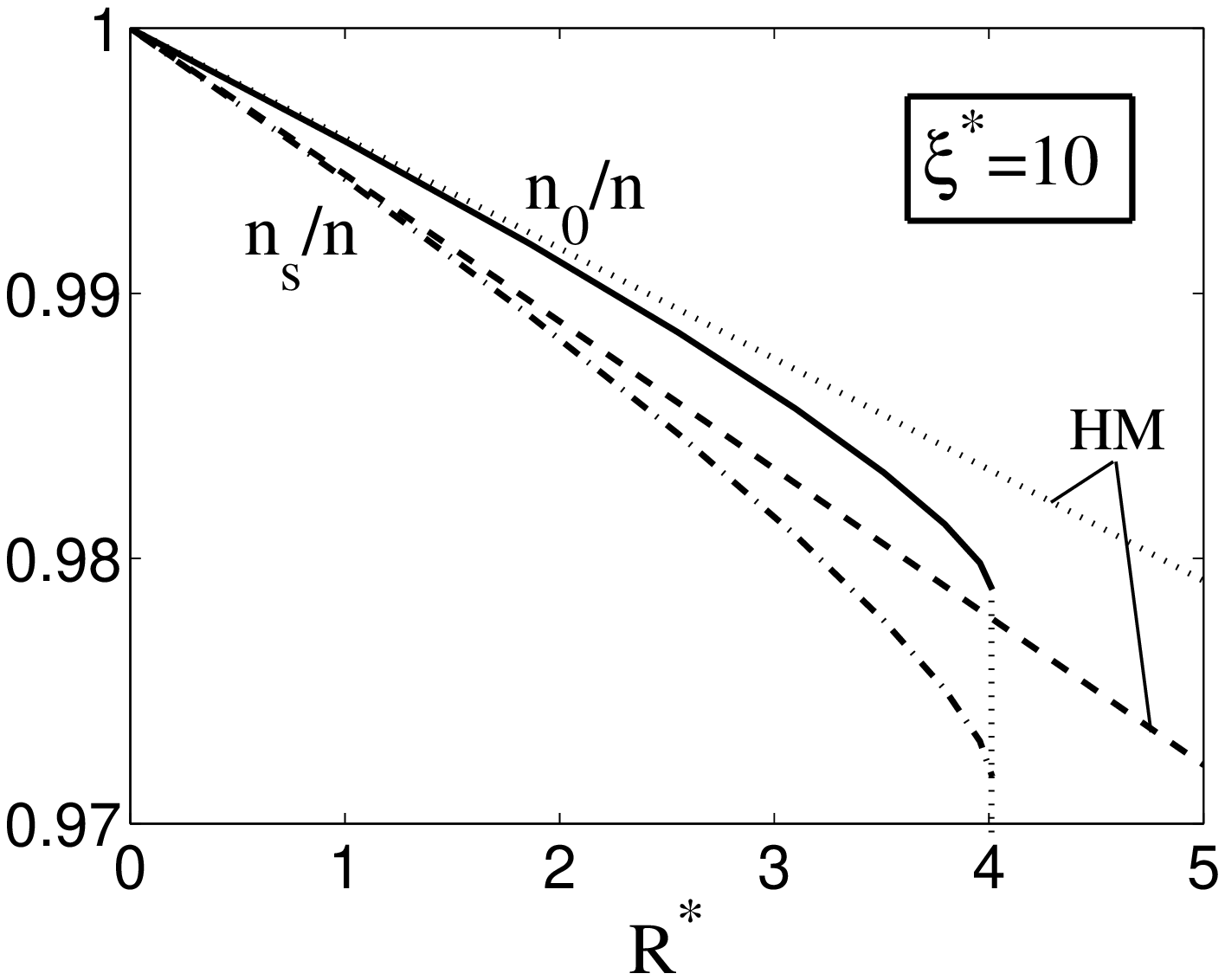}}
\caption{Same as for Fig.1 but for correlated disorder with $\xi^*=10$.
}
\label{fig3}
\end{figure}

\begin{figure}[t]
\scalebox{0.5}{
\includegraphics{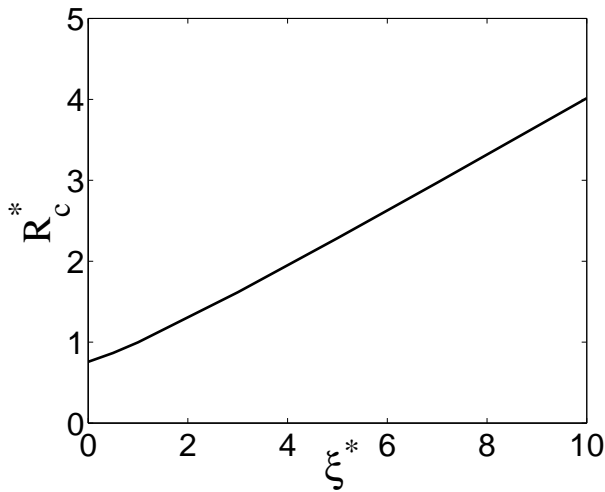}}
\caption{Critical value for the disorder intensity $R^*_c$ as a function of 
the correlation length $\xi^*$.}
\label{fig4}
\end{figure}
\section{Phase transition}
Equation (\ref{eqn}) is  solved numerically by an iterative procedure 
as a function 
of the reduced dimensionless parameters for the disorder strength  
$R^*=R m^{3/2}/\sqrt{4\pi^2gn/V}$ 
and the correlation length $\xi^*= \sqrt{4m g n/V} \xi $. We have looked 
for the stable solution that minimizes the total energy of the system. 
The homogeneous part of the 
condensate $n_0$ and the superfluid fraction $n_s$ 
are plotted in 
Figs. 1--3 as a function of the reduced disorder strength $R^*$ and are compared 
with the results obtained by HM. 
The transition is determined as the highest value $R^*=R^*_c$ for 
which the condensate fraction $n_0$ is non-zero so that spatial 
coherence is preserved over the entire space. 
This point corresponds to the situation where $n_0$ has an infinite derivative 
with respect to $R^*$. As a consequence of relation (\ref{hms}), 
also the superfluid density $n_s$  
has there an infinite derivative. 
For uncorrelated disorder i.e. $\xi^*=0$, we notice only a tiny difference 
with respect to the second-order theory, 
until we reach the critical value when the disordered part of the 
condensate is about 30 percent. For higher values of $\xi^*$ 
this difference becomes more pronounced, but a smaller fraction of the 
disordered part of the fluid is needed 
in order to achieve the transition. Fig.~4 shows that the 
critical value $R_c^*$ increases as a function of the coherence length 
which is understandable from the fact that for larger $\xi^*$ the high spatial 
frequency components of the disordered part of the condensate are smaller 
(cf. Eq.(\ref{eqn})).
%

\section{Conclusions}
The random phase approximation has been applied to the 
Gross-Pitaevskii equation in the presence of a random potential
in order to describe a strongly disordered Bose gas at zero temperature.  
This approximation goes beyond a previous second-order calculation and predicts 
a first-order phase transition from a superfluid phase to a non-superfluid phase. 
The critical value of the disorder intensity for this transition 
depends strongly on the correlation length. Nevertheless, our model 
fails to describe the properties of the non-superfluid phase.  
A possible explanation is that the assumption 
of a unique wave function for any particle
excludes  
the possibility of having fragmented condensates for 
strong disorder that could be necessary in a such a phase.
\section*{Acknowledgements}
This work was supported by the SFB/TR 12 of the German Research Foundation (DFG).
Furthermore, PN acknowledges support from the german AvH foundation, 
from the Belgian
FWO project
3E050202, and from 
Junior Fellowship F/05/011 of the KUL Research Council.
\end{document}